\documentclass[letterpaper,10pt]{extarticle}  % "extarticle" gives more global font size options. use 8pt with Montserrat and 10pt with EBGaramond
\usepackage{import}
\usepackage{local}
\usepackage[left=.65in,top=.65in,bottom=.65in,right=.65in]{geometry}
\usepackage{lipsum} % to generate text

\usepackage[version=3]{mhchem}
\usepackage{amsmath}
\usepackage{epstopdf}
\usepackage{hyperref}
\usepackage{array}
\usepackage{balance}
\usepackage{times,mathptmx}
\usepackage{graphicx} 
\usepackage{lastpage}

\newcommand{\etl}{\textit{et al.}}

\title{A Rigid Bender Study of the Bending Relaxation of \ce{H2O} 
	   and \ce{D2O} by Collisions with Ar}

\author{%
\textbf{Ricardo Manuel García-Vázquez \orcidlink{0000-0002-3786-8028},\textcolor{Accent}{\textsuperscript{1}} %
Lis\'an David Cabrera-Gonz\'alez \orcidlink{0000-0002-0991-9027},\textcolor{Accent}{\textsuperscript{2}} Otoniel Denis-Alpizar \orcidlink{0000-0002-0686-6927},\textcolor{Accent}{\textsuperscript{3,*}}\\
Thierry Stoecklin \orcidlink{0000-0002-1349-8106}\textcolor{Accent}{\textsuperscript{1,*}} }\\
\begin{small}\textcolor{Accent}{\textsuperscript{1}}UMR5255-CNRS, Universit\'e de Bordeaux, 351 cours de la lib\'eration , F-33405 Talence, France. \\ 
\textcolor{Accent}{\textsuperscript{2}}Department of Chemistry, The University of Manchester, Oxford
Road, Manchester M13 9PL, UK.\\
\textcolor{Accent}{\textsuperscript{3}}Grupo de Investigaci\'on en F\'isica Aplicada, Instituto de Ciencias Aplicadas,  Facultad de Ingenier\'ia, Universidad Aut\'onoma de Chile, Av. Pedro de Valdivia 425, Providencia, Santiago, Chile.\\
\textcolor{Accent}{\textsuperscript{*}}Correspondence: \textcolor{Accent}{otoniel.denis@uautonoma.cl,thierry.stoecklin@u-bordeaux.fr} \\ \end{small}
}

\date{14/11/2023}
\begin{document}
\maketitle
\thispagestyle{empty}

%%%%%%%%%%%%%%%%%%%%%%%%%%%%%%%%%%%%%%%%%%%%%%%%%%%%%%%%%%%%%%%%%%%%%
% Abstract
%%%%%%%%%%%%%%%%%%%%%%%%%%%%%%%%%%%%%%%%%%%%%%%%%%%%%%%%%%%%%%%%%%%%%
\section{Abstract}

\begin{doublespacing}
%\begin{linenumbers}

\noindent
\textbf{The bending relaxation of \ce{H2O} and \ce{D2O} by collisions with Ar is studied at the Close Coupling level. Two new 4D PES are developed for these two systems. They are tested by performing rigid rotor calculations as well as by computing the \ce{D2O-Ar} bound states. The results are compared with available theoretical and experimental data. Propensity rules for the dynamics are discussed and compared to those of \ce{H2O} colliding with  Ne or He. The bending relaxation cross sections and rates are then calculated for these two systems. The results are analysed and compared with available experimental data.}

%\begin{multicols*}{2} % if you want two columns
%%%%%%%%%%%%%%%%%%%%%%%%%%%%%%%%%%%%%%%%%%%%%%%%%%%%%%%%%%%%
\section*{Introduction}
\label{introduction}

Inelastic collisions between atoms and molecules play a fundamental role in atmospheric as well as interstellar chemistry. Up to very recently, detection was limited to the rotational spectrum of molecules but the ALMA interferometer as well as the James Web Space Telescope (JWST) opened the possibility of also detecting ro-vibrational transitions. This new possibility created the need to have access to vibrationally inelastic state collisional rates. Unfortunately, state-to-state resolved experimental rates, or cross sections, remain very difficult to measure and, when available, are limited to very few experimental points or to narrow temperature intervals. On the other hand, the progresses of computational capabilities allow to study more and more complicated systems and quantum scattering calculations including vibration became possible. The quantum studies in this domain are still scarce and the information gathered from the study of such a model system as \ce{H2O} colliding with Ar are quite valuable.

From the experimental point of view, the collisions of \ce{H2O} + noble gases, have taken on new relevance because of their relative simplicity  and their importance for atmospheric and ISM models. Among all the \ce{H2O} + Ar  has received a great deal of interest for many years, both on the theoretical and experimental sides. Most of these studies were devoted to the development of interaction potential and to the description of the bound states of the system\cite{hutson90,Lascola1991,bulski91,Chalasinki:91,tao:94_ar-h2o}.

The first attempt to experimentally study the dynamics of this system was made in 1975 by Bickes \etl \cite{elastic_exp_H2O}. In this work, the authors investigated the experimental elastic differential cross sections for several collision partners, of which one or both were polar molecules. Five years later, two new experimental works were done to measure the vibrational relaxation of \ce{D2O}(010) with molecular and atomic collision partners\cite{Sheffield1980,Miljanic1980}. However, the measured rates in these two studies  differ by a  factor of 70 \cite{Sheffield1980}. 

A new experimental study was carried out in 1989 by Zittel and Masturzo \cite{Zittel1989}who explored the vibrational relaxation of \ce{H2O} by collision with \ce{H2O} itself and with some rare gases in the [295,1000] K temperature interval. Due to the limitations of the experimental setup, no state-to-state rate constants were determined and they reported only upper limits for the bending relaxation rate of \ce{H2O}(010) + Ar. The first state-to-state measure for Ar + \ce{H2O} was done in 1999 by Chapman \etl \cite{chapman:99}. They studied the rotational excitation of \ce{H2O} from the fundamental \textit{para} and \textit{ortho} states at a collision energy of $(480\pm 90)$ cm$^{-1}$. They also made a comparison with close coupling (CC) scattering calculations performed using two \textit{ab-initio} Potential Energy Surfaces (PES) \cite{tao:94_ar-h2o,Szalewicz-pc}.

The first theoretical study of this system was done by  Ree and Shin\cite{Ree1990} in 1990. They used a semiclassical procedure to investigate the vibrational relaxation of \ce{H2O} and \ce{D2O} by collision with Ar. The collisional rates were not calculated and only transition probabilities were reported which agreed with experiment. Recently, in 2017, Ndengu\'e \etl\cite{Ndengue2017} presented a theoretical study of the rotational excitation using the MultiConfiguration Time-Dependent Hartree (MCTDH) approach as well as CC calculations. Very recently, a full dimensional studies of Ar + \ce{H2O} collision was reported by the Guo's team \cite{Guo_ar_2022,Guo_ar2_2022} in 2022. They used a full dimensional PES\cite{Liu2022} and studied the vibrational relaxation of the system using the CC method and tested the validity of the Rigid Rotor Approximation for $\nu_2 = 0,\nu_2 = 1$ levels. While this latter study demonstrated that  full dimensional calculations are possible for this system, the coupling between bending and stretching is quite small in the $ \nu_2 = 1 $ state making possible to carry out  scattering calculations using the Rigid-Bender approximation\cite{Stoecklin-rb-2019} in order to compare our results with the experimental works available as was already done successfully for the collisions of  \ce{H2O} with He \cite{Stoecklin-H2o-He-2021} and H \cite{lisan-h-h2o}. The other important objective of the present work is to study the inelastic collision of Ar with \ce{D2O} and to compare the two systems as the only study investigating these effects was done using a  semi-classical approach\cite{Ree1990}. 

To this aim  two 4D PES for \ce{Ar-H2O} and \ce{Ar-D2O}, including an explicit dependence on the bending angle are developed. They are tested at the rigid rotor level by performing  scattering calculations as well as computing  the \ce{D2O-Ar} bound states and comparing the results with available theoretical and experimental data. The Dynamics propensity rules are compared for these two systems and to those obtained previously for the rotationally inelastic collisions of \ce{H2O} with  Ne or He. The bending relaxation of \ce{H2O} and \ce{D2O} by collisions with Ar are then studied at the Close Coupling level taking fully into account the coupling between bending and rotation within the RB-CC method. The results obtained for these two systems are compared between each other and with their experimental estimates.

The paper is organized as follows: In section \nameref{Meth},  the methods used to produce the two new 4D PESs, and to perform the scattering calculations are briefly reminded. The results are then analysed and  compared with experimental and theoretical data available in section \nameref{CalRes} while a few conclusions of the present study are given in the \nameref{Concl} section.

\section*{Method} 
\label{Meth}
\begin{figure}
	\centering
	\includegraphics[width=100mm]{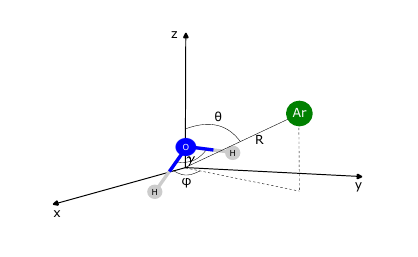} 
	\caption{System of coordinates used to describe the interaction between \ce{H2O} and Ar}
	\label{fig1}
\end{figure}

The system of coordinates origin is fixed at the \ce{H2O} center of mass and the \ce{H2O} molecule is lying in the ZX plane with the Z axis bisecting the \ce{H2O} bending angle as shown in Figure \ref{fig1}were the system Jacobi coordinates used both for the PES and the dynamics is represented. 
$\gamma$ is the \ce{H2O} bending angle while $  \overrightarrow{R} $  is the \ce{H2O-Ar} intermolecular vector which spherical coordinates are $ (R,\theta,\varphi)$.  Within the rigid bender approximation the  \ce{OH}  distance was fixed at its equilibrium value of 0.957 \AA\cite{valiron:08r12,tao:94_ar-h2o}.

\label{Fig:1}

\subsection*{Potential Energy Surface}

The electron correlation CCSD(T) level of theory together with an aug-cc-pVQZ basis set \cite{dunning:89}  as well as
a supplementary set of bond orbitals\cite{cybulski:99} were used to compute a grid of  14896  \textit{ab initio} geometries of the \ce{Ar-H2O} system including the \ce{H2O} bending motion and using the \texttt{MOLPRO} package\cite{molpro} . The basis set superposition error was corrected using the counterpoise method\cite{boys:70}. The grid includes 28 values of R distances ranging from 1.5 \AA \!  up to 12 \AA, 7 \ce{H2O} bending angles taken in the  [70$^\circ, 140^\circ$] interval and including its equilibrium value (104.41 $^\circ$). For each value of  R and $ \gamma $, 76
angular configurations were obtained by combining 19 values of  $ \theta  $ in the $[0^\circ, 180^\circ]$ interval and 4 values of $ \varphi $ ranging from $0^\circ$ to  90$^\circ$.

\subsubsection*{Fitting Procedure}
The grid was divided into two parts, the short range (SR) for $R< 7.0$ \AA \,  and the long range (LR) for $R \geq 7.0$ \AA. The SR and LR grids were  independently expanded in the Green basis set\cite{green:76} using a weighted least-square method
\begin{equation}
	V(\gamma,R,\theta,\varphi) = \sum_{l=0}^{l_{max}} \sum_{m=0}^{min(l,m_{max})}\nu_{l,m}(\gamma,R) \bar{P}_{l}^{m}(cos(\theta))cos(m\varphi)	\label{eq1}
\end{equation}   
with $(l_{max},m_{max})$ respectively equal to $ (17,6) $ and $(6,2)$ for the SR and LR grids  while only even values of $m$ were considered due to the $C_{2v}$ symmetry of the H$_2$O molecule. The $\nu_{l,m}$ coefficients were then fitted using the Reproducing Kernel of the Hilbert Space (RKHS) procedure \cite{ho:96} as:
\begin{equation}
	\nu_{l,m}(\gamma,R) = \sum_{i=1}^{N_R \cdot N_{\gamma}} \alpha_{lmi}q_{R}^{(2,5)}(R,R_i)q_{\gamma}^{(2)}(z(\gamma),z(\gamma_i)) \label{eq2}
\end{equation}
where $N_R$ and $N_\gamma$ are the number of values of $R$ and $\gamma$ while $q_{R}^{(2,5)}$ and $q_{z(\gamma)}^{(2)}$ are the  usual one-dimensional kernel functions \cite{ho:96}:
\begin{align}
	q_{R}^{(2,5)}(R,R_i)  &= \dfrac{2}{21 R_{>}^6} - \dfrac{R_{>}}{14 R_{>}^7}\\
	q_{\gamma}^{(2)}(z(\gamma),z(\gamma_i)) &= 1 + z_{<}z_{>} + 2(z_{<})^2z_{>}\left(1-\dfrac{z_{<}}{3z_{>}}\right),
\end{align}
with $ z(\gamma) = \dfrac{1-\cos(\gamma)}{2}$. $R_{>}$ and $R_{<}$ are respectively the higher and lower values of $R$ and $R_i$ while $z_{>}$ and $z_{<}$ are respectively the higher and lower values of $z$ and $z_i$    . The final four-dimensional potential energy surface is obtained as  a combination of the LR and SR parts using the switching function $ S(R) $\\
\begin{align}
	V(\gamma,R,\theta,\varphi) &= S(R)V^{(SR)}(\gamma,R,\theta,\varphi) + (1-S(R))V^{(LR)}(\gamma,R,\theta,\varphi) \\
	S(R) &= \frac{\mathrm{erfc}(A_0(R-R_0))}{2}
\end{align}
where $V^{(SR)}$ and $V^{(LR)}$ are the SR  and LR parts of the PES.  The fitting parameters $A_0= 3$\AA{}$^{-1}$ and $R_0 = 7$\AA{} where obtained by the trial and error method.
\subsubsection*{\ce{Ar-D2O} PES} 
We use the same coordinate system as for \ce{H2O} + Ar, but the frame origin is now placed at the \ce{D2O} center of mass. The experimental equilibrium geometries of \ce{H2O} and \ce{D2O} are very close, so we take them as equal. In order to obtain a new PES for \ce{D2O} + Ar, a dense grid of points is first generated using the  PES developed for \ce{H2O} + Ar and making the change of coordinates associated with the new center of mass. The set of formula necessary to make the change of coordinates  was given in a previous paper \cite{Garcia-Vazquez_2023} while  in the present  case, formula (8) of that  paper is now a function of the bending angle.   This new grid is composed of 34 values of the \ce{Ar-D2O} distances in the ($2.3\le R \le 12 $ \AA{}) interval while the grid of bending and polar angles is identical to the one used for \ce{Ar-H2O}. We use the same procedure explained before for the \ce{Ar-H2O} PES to generate a new fit for the \ce{Ar-D2O} system.

\subsection*{Dynamics}
The Close Coupling calculations were carried out separately for  each of the \textit{ortho-}  and \textit{para-} forms of \ce{H2O}/\ce{D2O} using our new PES and the \texttt{Newmat} code\cite{Stoecklin-rb-2019}, following the same methodology  than in our  previous works \cite{Stoecklin-rb-2019,Stoecklin-H2o-He-2021,lisan-h-h2o}. We performed both rigid rotor (RAST-CC ) and rigid bender (RB-CC ) calculations. The basis set describing the \ce{H2O}/\ce{D2O} molecules included 10 values of the rotational quantum number $j$ for the RAST-CC calculations while the RB-CC  one  included 10 values of the rotational quantum number $j$  inside each of the two first vibrational bending levels. Collision energy is ranging respectively in the [0.01,4000] cm$^{-1}$ and [0.01,2000] cm$^{-1}$   intervals for the RAST-CC and RB-CC calculations. The radial parts of the RAST-CC and RB-CC scattering wave functions were propagated up to 40 $a_0$ using the log derivative method \cite{manolopoulos:88thesis} and the convergence was checked as a function of the propagator step size. Due to the large relative mass of the systems, it was necessary to include partial waves up to $J=200$  to reach a 10$^{-2}$ relative convergence criterion of the state-selected quenching cross section as a function of  $J$. For comparison, in our previous studies dedicated to  \ce{H-H2O} and \ce{He-H2O}, $J$=65 and $J$=125 respectively were sufficient to achieve a  10$^{-3}$ relative  convergence.

\section*{Calculations and Results}\label{CalRes} 
\subsection*{\ce{Ar-H2O} PES}
The global minimum  of the present \ce{Ar-H2O} PES was searched using several two-dimensional downhill simplex procedures\cite{num_recipes_f77} and found to be $-203.19$ cm$^{-1}$  for the ( $\gamma = 70^\circ$, $R = 3.56 \AA$, $\theta = 180^\circ$)  geometry of this complex.  It is very close to the \textit{ab initio} minimum ($-202.25$ cm$^{-1}$) of our grid located at  ($\gamma = 70^\circ$, $R = 3.6 \AA$, $\theta = 180^\circ$ ) . The root mean square deviation (RMSD) for negative energies ($E\leq 0$ cm$^{-1}$)  is 5.44$\times10^{-3}$ cm$^{-1}$ while the RMSD for positive energies ($0$ cm$^{-1}$ $<E\leq 5000$ cm$^{-1}$) is also quite satisfactory: 4.32$\times10^{-3}$ cm$^{-1}$. 

\begin{figure}
	\centering
	\includegraphics[width=8.4cm]{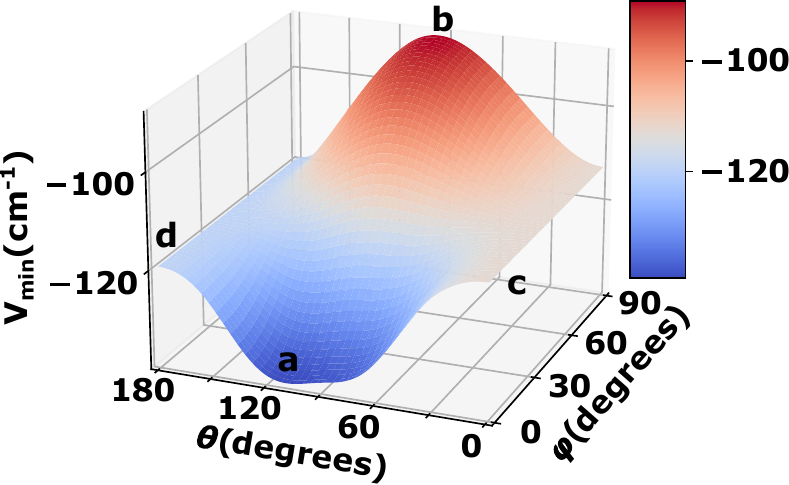}
	\caption{ $V_{\gamma_e,R_{min}}(\theta,\varphi)$ PES obtained by minimizing the energy as a function of R for the equilibrium bending angle ($\gamma_e$)  and a given ($\theta,\varphi$) pair. $a-d$ are stationary points where $a$ is the minimum at $\gamma=\gamma_e$. }
	\label{minpes1}
\end{figure}

The energy minimum of the complex ($-139.37$ cm$^{-1}$) obtained for the \ce{H2O} equilibrium bending angle  ($\gamma_e$) is associated with a nearly linear hydrogen(deuterium)-bonded configuration, at ($R = 3.67$ \AA$ $, $\theta = 109.38^{\circ},\varphi = 0^{\circ}$). 
Figure \ref{minpes1} shows a minimized PES $V_{\gamma_e,R_{min}}(\theta,\varphi)$ obtained by minimizing the energy with respect to $R$ for each ($\theta,\varphi$) pair where several stationary points are observed, including the minimum $a$ at $\gamma_e$ . Table \ref{pes} presents  a detailed comparison between  our stationary points and those of previous works which are found to be in very good agreement \citep{h2o_ar_hou_2016,tao_h2o_ar_1994,cohen_h2o_ar_1993,Liu2022}.

\begin{table*}
	\caption{Comparison between the present Ar-\ce{H2O} stationary points obtained for the \ce{H2O} equilibrium bending angle and those obtained in previous works. The distances, angles and energies are given in \AA, degrees and cm$^{-1}$, respectively. The labels of the stationary points matches the ones on Fig. \ref{minpes1}.}
	\label{pes}
	\begin{tabular}{clcccc}
		\hline
		\hline
		Stationary point 	&PES & $R$		&	$\theta$		&	$\varphi$  &  $Energy$	\\
		\hline
		& This work (CCSD(T)/aug-cc-pVQZ+bf) & 3.67 & 109.38 & 0 & $-139.37$ \\
		& Liu 2022 \cite{Liu2022} (FC-CCSD(T)-F12a/aug-cc-pVQZ+bf) & 3.66 & 108.71 & 0 & $-140.63$ \\
		$a$ & Hou 2016 \cite{h2o_ar_hou_2016} (CCSD(T)/aug-cc-pVQZ+bf) & 3.63 & 100.05 & 0 & $-139.53$ \\
		&  Tao 1994 \cite{tao_h2o_ar_1994} (MP4/basis+bf) & 3.60 & 105.00 & 0 & $-130.20$ \\
		&  Cohen 1993 \cite{cohen_h2o_ar_1993} (empirical-AW2) & 3.64 & 105.70 & 0 & $-142.98$ \\
		\hline
		& This work &3.69 & 96 & 90 & $-89.03$ \\
		& Liu 2022 & 3.68 & 90 & 90 & $-91.00$ \\
		$b$ & Hou 2016 & 3.69 & 90 & 90 & $-89.85$ \\
		& Tao 1994 & 3.76 & 90 & 90 & $-77.60$ \\
		& Cohen 1993 & 3.67 & 90 & 90 & $-89.70$ \\
		\hline
		& This work & 3.57 & 0 & -- & $-112.62$ \\
		& Liu 2022 & 3.56 & 0 & -- & $-115.02$ \\
		$c$  & Hou 2016 & 3.57 & 0 & -- & $-113.48$ \\
		& Tao 1994 & 3.62 & 0 & -- & $-103.60$ \\
		& Cohen 1993 & 3.52 & 0 & -- & $-125.8$ \\
		\hline
		& This work & 3.66 & 180 & -- & $-119.40$ \\
		&  Liu 2022 & 3.66 & 180 & -- & $-120.91$ \\
		$d$  &Hou 2016 & 3.66 & 180 & -- & $-118.95$ \\
		& Tao 1994 & 3.71 & 180 & -- & $-107.60$ \\
		& Cohen 1993 & 3.70 & 180 & -- & $-116.69$ \\
		\hline
		\hline									
	\end{tabular}
\end{table*}

\begin{figure*}
	\begin{center}
		\includegraphics[width=\textwidth]{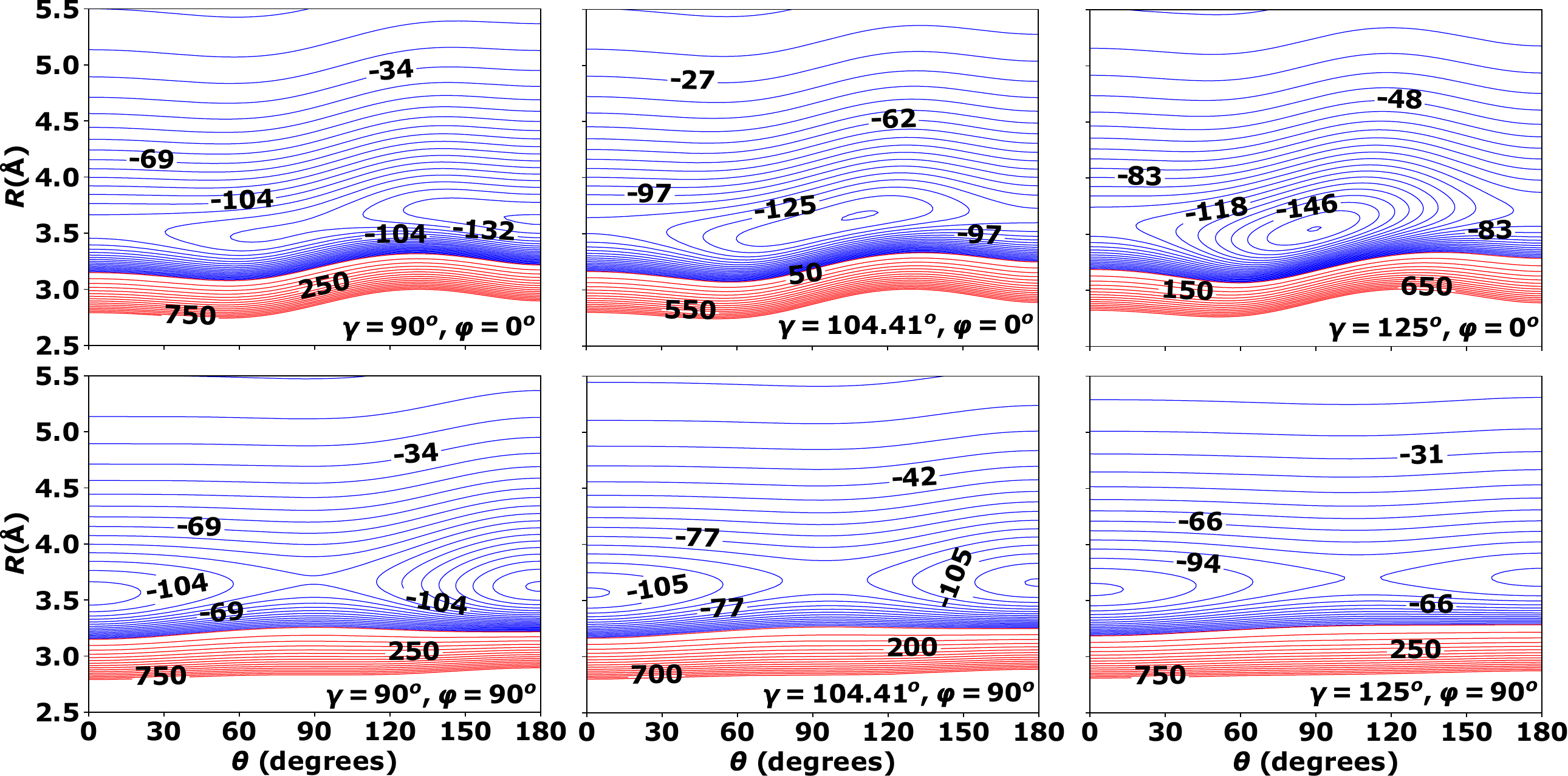}
		\caption{Contour plots of the \ce{H2O} + Ar complex for several values of $\gamma$ and $\varphi$. Negative (blue) and positive (red) contour plots are reported in steps of 7 cm$^{-1}$ and 50 cm$^{-1}$, respectively.}
		\label{3dpes}
	\end{center}
\end{figure*}

Another representation of  the PES is shown in Fig. \ref{3dpes} where contour plots of $V(\gamma,R,\theta,\varphi)$ are presented for several fixed values of $\gamma$ and $\varphi$. The lower energies are seen to be associated with planar configurations of the system ($\varphi = 0^\circ$) while the PES appears to change significantly when the bending angle is varied. Additionally, the $\theta$ value corresponding to the minimum is found to decreases from 180$^\circ$ up to $\sim$88$^\circ$ when the bending angle increases. 

\subsection*{\ce{Ar-D2O} PES}
The minimum energy value ($-139.37$ cm$^{-1}$) obtained for  the \ce{D2O} equilibrium bending angle ( $R = 3.63$\AA{}, $\theta = 108^{\circ}$, $\varphi = 0 ^{\circ}$) is quite close to the one of the \ce{Ar-H2O} PES given in table \ref{pes} and is also in good agreement with the one reported by Wang \etl \cite{Wang2015} ($R=3.65$\AA{}, $\theta = 105.75 ^{\circ}$, $\varphi = 0 ^{\circ}$, $E = -139.2$ cm$^{-1}$)\footnote{the value for $\theta$ angle given in this reference (Ref \cite{Wang2015}), $\theta = 74.25^{\circ}$, correspond to the supplementary angle of our $\theta$ angle}. The accuracy of our  PES can also be favourably appreciated by looking at the RMSD values which is   $7.94\times 10^{-3} $ cm$^{-1}$ for negative energies and $1.18\times 10^{-4}$ cm$^{-1}$ for  positive one. 

\begin{figure}
	\centering
	\includegraphics[width=\textwidth]{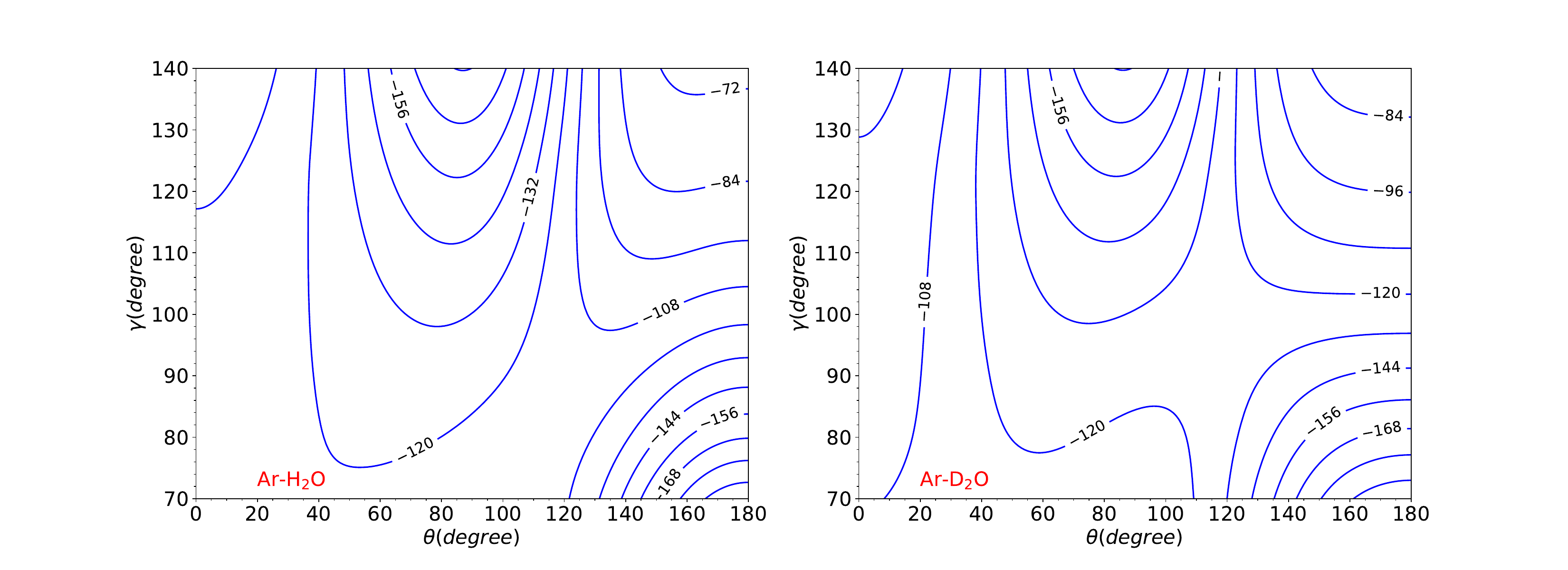}
	\caption{ Contour plots of the \ce{Ar-H2O} (upper panel) and \ce{Ar-D2O} (bottom panel) PES for $R = 3.5$ \AA{} and $\theta = 0 ^{\circ}$. }
	\label{gammtht}
\end{figure}

In Fig. \ref{gammtht} we compare contour plots of the two systems PES for $R = 3.5$ \AA{} and $\varphi = 0^{\circ}$. The upper and bottom panels are respectively used for \ce{Ar-H2O} and \ce{Ar-D2O}. The shape of the barrier is seen to change from one system to the other which is expected  to produce differences between the dynamics of the two systems.  

\subsection*{\ce{Ar-D2O} Complex Bound States}
We calculated the \ce{Ar-D2O} bound states and compare them with the calculation reported by Wang \etl \cite{Wang2015} in order to check the accuracy of our \ce{Ar-D2O} PES. The calculations were performed at the rigid rotor (RR) level using the same approach as in previous work \cite{Stoecklin-H2o-He-2021}. The results shown in Table \ref{bound}  present a very good agreement with the data reported by Wang \etl \cite{Wang2015}. The average  difference percentage are respectively 0.64 \% and 0.84 \%  for the \textit{ortho} and \textit{para} states, suggesting that our \ce{Ar-D2O} PES is accurate enough to be used in scattering calculations.

\begin{table*}
	\caption{Bound states of the \ce{Ar-D2O} complex. The energies are in cm$^{-1}$ and the values in parenthesis are those reported in Ref. \citenum{Wang2015}.}
	\label{bound}
	\begin{tabular}{cccccccc}
		\hline
		\hline
		Assignment &	$J=0$		&	$J=1$		&	$J=2$	& Assignment &	$J=0$		&	$J=1$		&	$J=2$\\
		\hline
		\ce{Ar-oD2O} ($\nu_2 = 0$)	&	&	&& \ce{Ar-pD2O}($\nu_2 = 0$) & & &\\
		$\Sigma(0_{00})^e$	&	$-95.9232$ & $-95.7375$	& $-95.3662$ & 
		$\Sigma(1_{01})^e$  & $-90.1735$	   & $-89.9922$	& $-89.6299$	\\
		&	($-95.2999$) & ($-95.1157$) & ($-94.7473$) & &($-89.5692$)	  &($-89.3891$)	&($-89.0291$) \\
		$\Pi(1_{11})^e$		&				& $-76.4708$	& $-76.1594$ &  $\Pi(1_{01})^f$		&			 & $-79.0979$	& $-78.7313$	\\	
		&	    	   & ($-75.9139$) & $-75.6081$	 & &			   & ($-78.5907$)	& ($-78.2300$) \\
		$\Pi(1_{11})^f$		&			   & $-76.4397$	& $-76.0688$   & $\Pi(1_{01})^e$		&			   & $-79.0924$	& $-78.7149$	 \\
		&			   & ($-75.881$)  & ($-75.5128$)  & &			   & ($-78.5854$)	& ($-78.2140$) \\
		$\Sigma (1_{11})^e$ &	$-74.6936$   & $-74.4764$	& $-74.0449$	 & $\Pi(1_{10})^f$		&			   & $-75.7615$	& $-75.3950$	\\
		&  ($-74.3293$)  & ($-74.1115$) & ($-73.6795$)  & &			   & ($-75.1121$)	& ($-74.7443$) \\
		$\Sigma (2_{02})^e$ &	$-64.3926$   & $-64.2436$	& $-63.9446$	&$\Pi(1_{10})^e$		&			   & $-75.7571$	& $-75.3819$	 \\
		&	($-63.7309$) & ($-63.5861$)	& ($-63.2953$) & &			   & ($-75.1077$)	& ($-74.7311$)\\
		$\Pi (2_{02})^e$    &			   & $-60.8061$	& $-60.6248$ 	& $\Sigma (1_{10})^f$ & $-61.3964$	   & $-61.2099$	& $-60.8369$	\\
		&			   & ($-60.3039$)	& ($-60.0794$) & & ($-60.8469$)   & ($-60.6615$)	& ($-60.2909$) \\
		$\Pi (2_{02})^f$    &			   & $-60.6255$   & $-60.2682$ 	& $\Sigma (2_{12})^e$ & $-59.3582$	   & $-59.1984$	& $-58.8784$	 \\
		&			   & ($-60.2019$)	& ($-59.8462$) &&($-60.1083$)	   & ($-59.9452$)	& ($-59.6189$)  \\
		$1\Sigma(0_{00})^e$	&	$-60.5615$   & $-60.1751$   & $-59.5885$  	&$\Pi (2_{12})^e$    &			   & $-53.944$	& $-53.7472$		\\
		&	($-59.8961$) & ($-59.5864$) & ($-59.0416$) & &			   &($-53.2424$)	&($-52.9334$)	\\
		&&&&$\Pi (2_{12})^f$    &			   &$-53.8032$	&$-53.4499$	\\
		&&&&&			   &($-53.2206$)	&($-52.8692$)	\\
		
		\hline
		\hline									
	\end{tabular}
\end{table*}

\subsection*{Rotational Excitation} 
\begin{figure}
	\centering
	\includegraphics[width=\textwidth]{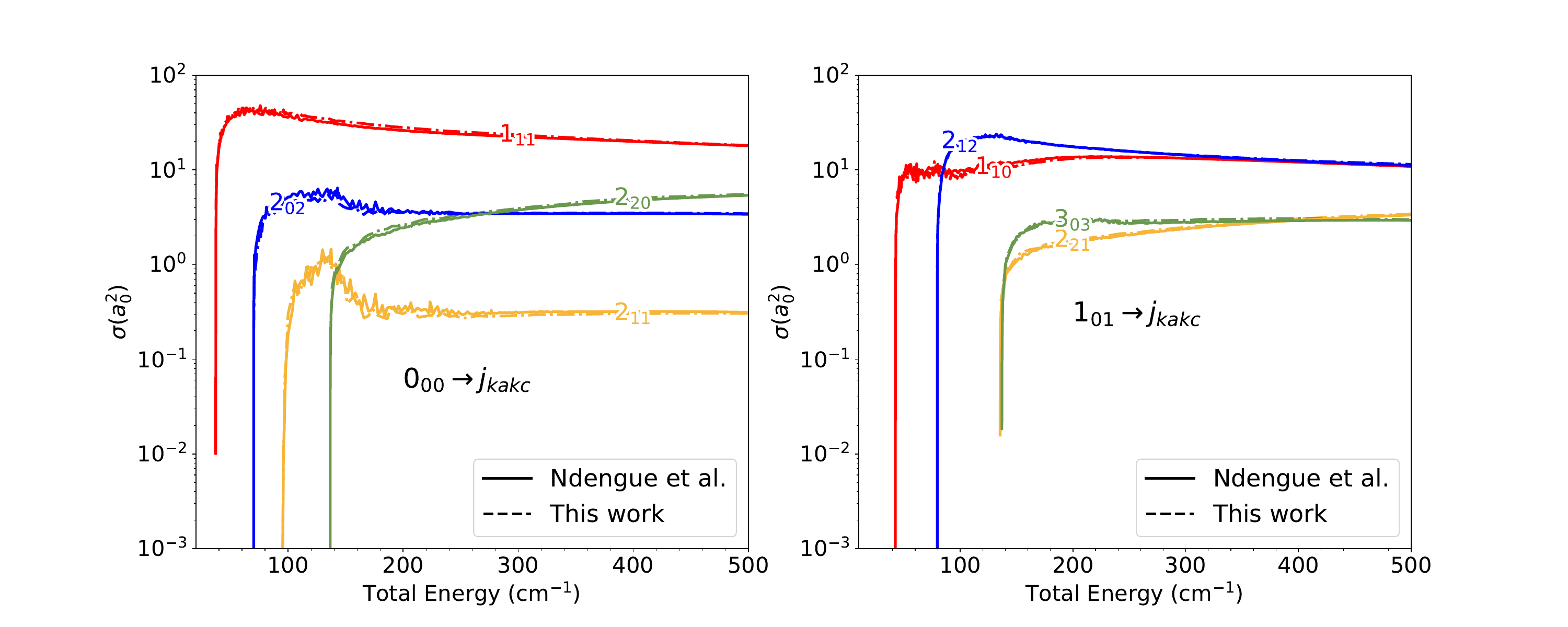} 
	\caption{Rotational Excitation of the fundamental \textit{para/ortho} levels of the water molecule by collision with Ar. Left and right panels present the excitation of the \textit{para} and \textit{ortho} states respectively. Solid lines correspond to the values obtained in this work while dashed lines are associated with the CC results of Ndengu\'e \etl \cite{Ndengue2017}.}
	\label{excit}
\end{figure}

We first computed rigid rotor excitation cross sections for a total energy of 480 cm$^{-1}$  in order to compare our results with  those of previous theoretical \cite{Ndengue2017} and experimental \cite{chapman:99} works available. The rotational states of the water molecule were labelled using the standard $j_{k_ak_c}$ notation. Here, $j$ represents the rotational angular momentum of the \ce{H2O} molecule, while $k_a$ and $k_c$ represent its projections over the $A$ and $C$ molecular axis, respectively\cite{zare:88}. Our results are compared  in Fig. \ref{excit} with the RR Close Coupling (CC) results of Ndengu\'e \etl \cite{Ndengue2017}. Although the potential energy surfaces are different, the results are in excellent agreement for all the  transitions. They  surprisingly differ by a factor 2 with the one reported by Liu \etl \cite{Guo_ar2_2022} for the transitions from $0_{00}$ to $2_{02}$ and $2_{20}$ and we cannot see any reason for this discrepancy.  

\begin{figure*}
	\begin{center} 
		\includegraphics[width=17.4cm]{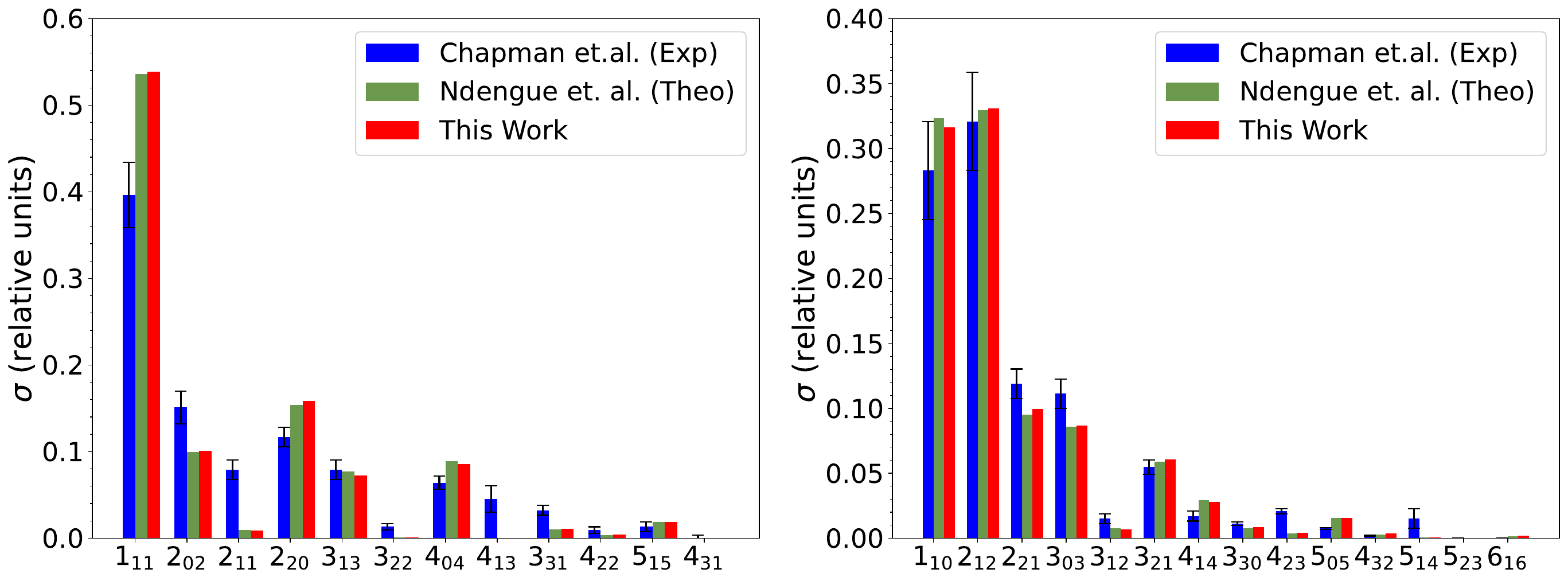} 
		\caption{Rotational Excitation of water by collision with Ar atoms for a total energy of 480 $cm^{-1}$. The left and right panels present respectively the cross sections for the excitation of the fundamental  \textit{para} $0_{00}$ and  \textit{ortho} $1_{01}$ levels. The experimental values of  Chapman \etl \cite{chapman:99} are shown in blue with error bars while green and red colors are used for the theoretical results obtained by Ref. \citenum{Ndengue2017} and by us respectively.}
		\label{comp-exp-480}
	\end{center}
\end{figure*}

Our results are also in very good agreement with the experimental results of  Chapman \etl \cite{chapman:99} as shown in Fig. \ref{comp-exp-480}.  The largest difference between theory and experiment  is seen to occur for the \textit{para} $2_{11}$ final level. In this case the calculations underestimate the reported experimental values by almost one order of magnitude. A similar difference was found by  Chapman \etl \cite{chapman:99} when they compared his results  with the theoretical one of  Tao and Klemperer \cite{tao:94_ar-h2o} based on AW2 PES\cite{tao:94_ar-h2o} and a SAPT PES\cite{Szalewicz-pc}.

In general, both theory and experiment predict strong alignment effects for both \textit{para} ($\sigma_{2_{02}\leftarrow 0_{00}}>\sigma_{2_{11}\leftarrow 0_{00}}<\sigma_{2_{20}\leftarrow0_{00}}$) and \textit{ortho} water ($\sigma_{3_{03}\leftarrow 1_{01}}>\sigma_{3_{12}\leftarrow 1_{01}}<\sigma_{3_{21}\leftarrow1_{01}}$). This behaviour was analysed by  Chapman \etl \cite{chapman:99} to result from the following propensity rule: When multiple final states exist for the same rotational angular momentum $j$ of the water molecule, the larger cross sections are obtained for $\Delta k_a\approx\Delta j$ or $\Delta k_c\approx\Delta j$ and the smaller cross sections for $\Delta k_a\approx\Delta k_c$. The  \textit{ortho/para} cross-sections ratio predicted by our calculations $\sigma_{1_{01}}/\sigma_{0_{00}} = 0.96$ is also in good agreement with the experimental one $\sigma_{1_{01}}/\sigma_{0_{00}} = 0.98$.

\subsection*{Rotational Relaxation of \ce{H2O} and \ce{D2O }: Propensity Rules}
\begin{figure*}
	\begin{center}
		\includegraphics[width=17.4cm]{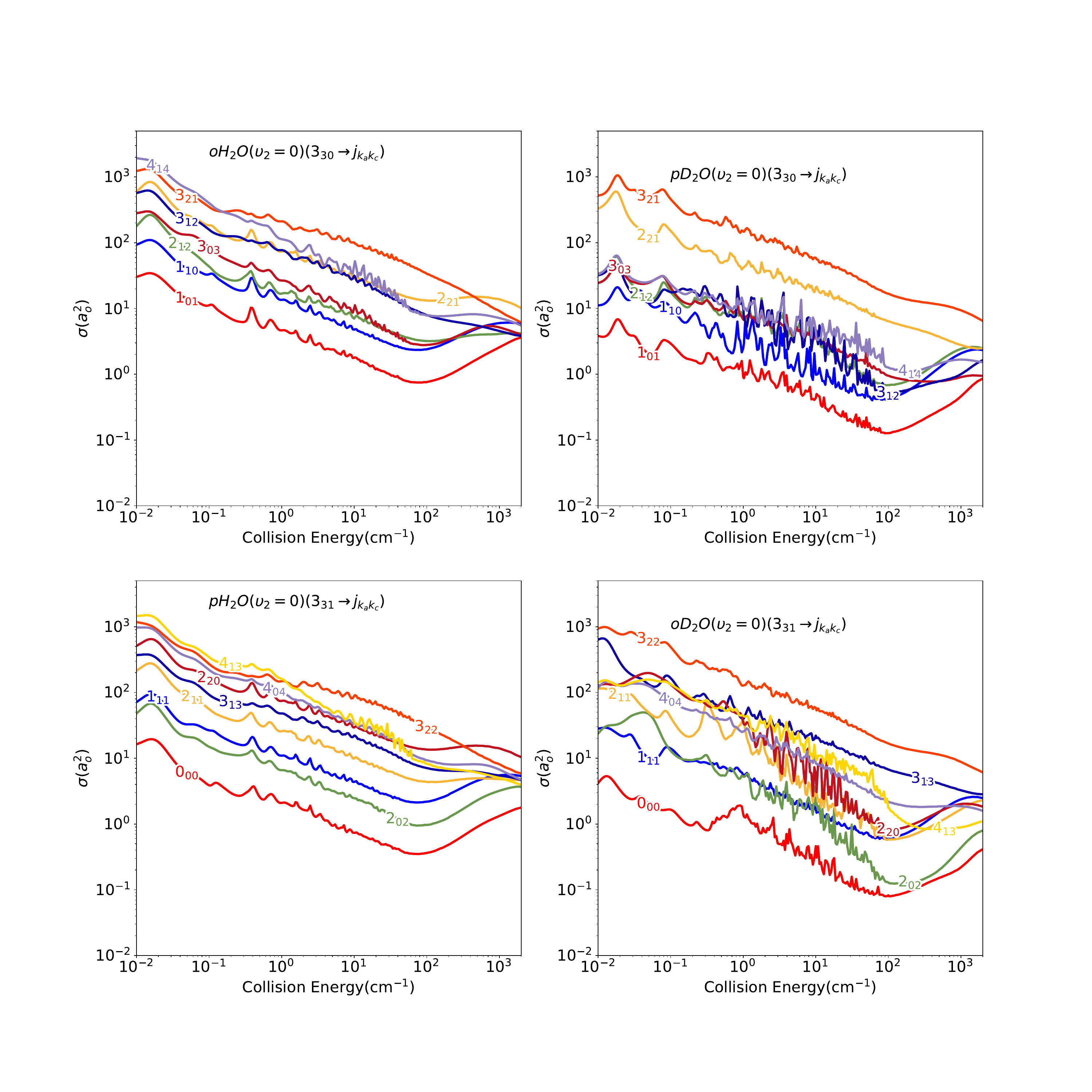} 
		\caption{Cross sections for the rotational relaxation of water and \ce{D2O} inside the fundamental bending level by collision with Ar. Top and bottom panels present respectively the relaxation of the $3_{30}$ and $3_{31}$  levels.  Left and right panels are used for \ce{H2O} and \ce{D2O} respectively.}
		\label{rr330}
	\end{center} 
\end{figure*}

State to state cross sections for pure rotational relaxation of the $3_{30}$ and $3_{31}$ levels of both molecules are compared in Fig. \ref{rr330}. The \ce{D2O} relaxation curves exhibit strong resonances which are seen to be weaker in the case of  \ce{H2O} as a result of the larger gap between successive levels of \ce{H2O} compared to \ce{D2O}. For both systems, only the relaxation cross sections towards the state the closer in energy with the initial state decrease monotonously as a function of collision energy while the others  as usual first decrease until they reach a minimum at collision energies roughly corresponding to the well depth of the PES and then start to increase again.

Propensity rules for the rotational relaxation of Ar + \ce{H2O} were proposed by Liu \etl \cite{Guo_ar2_2022} who suggested that the dominant final states follow the $\Delta j = \Delta k_c = -1$ rule. We also recently  obtained more general propensity rules for  the Ne + \ce{H2O} system \cite{Garcia-Vazquez_2023} and  wondered if they could hold  for the Ar + \ce{H2O} system . We then analysed all the relaxation cross sections involving   $j=0,1,2,3,4$ states of \ce{H2O} by collision with  Ar. We indeed verified that the $3_{03}$ and $3_{13}$ levels discussed by  Liu \etl \cite{Guo_ar2_2022} follow the propensity rule  $\Delta j = \Delta k_c = -1$ which is a peculiar case of the general rules proposed in our previous work \cite{Garcia-Vazquez_2023} when there is not any possible transition associated with $\Delta j=0$ and $\Delta k=0$ where k is the Z space fixed projection of the \ce{H2O} rotational angular momentum. In those cases we found that the dominant transition is always associated with the smallest possible  change in $j$ and $k$ which can be written $\Delta j = \pm 1$, $\Delta k = \pm 2$ for the  $3_{03}$ and $3_{13}$ levels. We verified that this rule also holds for transition starting from the $4_{04}$ and $4_{14}$ levels. 

We find more generally that the rules obtained for Ne + \ce{H2O} and which are also valid for He + \ce{H2O} stands for Ar + \ce{H2O} in almost  the entire energy interval considered in the present work . The transitions starting for example from  the $3_{21}$ and $3_{22}$  states follow the same order by magnitude that the cross sections for Ne + \ce{H2O} presented in Fig. 8 of  Garc\'ia-V\'azquez \etl \cite{Garcia-Vazquez_2023} with just an  interchange in the order of the $1_{01}$ and $2_{21}$ curves for the transitions starting from the $3_{21}$ level.

At very low collision energy [0.01,1] cm$^{-1}$ as illustrated in  Fig. \ref{rr330}, the transitions associated with the smallest energy gap may  however becomes dominant. At intermediate energies smaller than the potential well depth, the dominant transition  follows the $\Delta j=0, \,\Delta k=0$ rules while for higher energies they fulfil the  $|\Delta k_a|=|\Delta k_c| = 1$  rule, as in the case of Ne + \ce{H2O}.

We conversely did not find any obvious rule for the collision with D$_2$O, except for the most excited  level inside any $j$ multiplet which follows the rules obtained for the collisions with \ce{H2O}. However, the state to state cross sections  in Ne + \ce{D2O} and Ar + \ce{D2O} seem to follow the same ordering suggesting that some more complicated rules associated with the smaller energy gaps in \ce{D2O} are at play and still need to be inferred.

\subsection*{Vibrational Relaxation}
\subsubsection*{General Trends}
\begin{figure*}
	\centering
	\includegraphics[width=17.4cm]{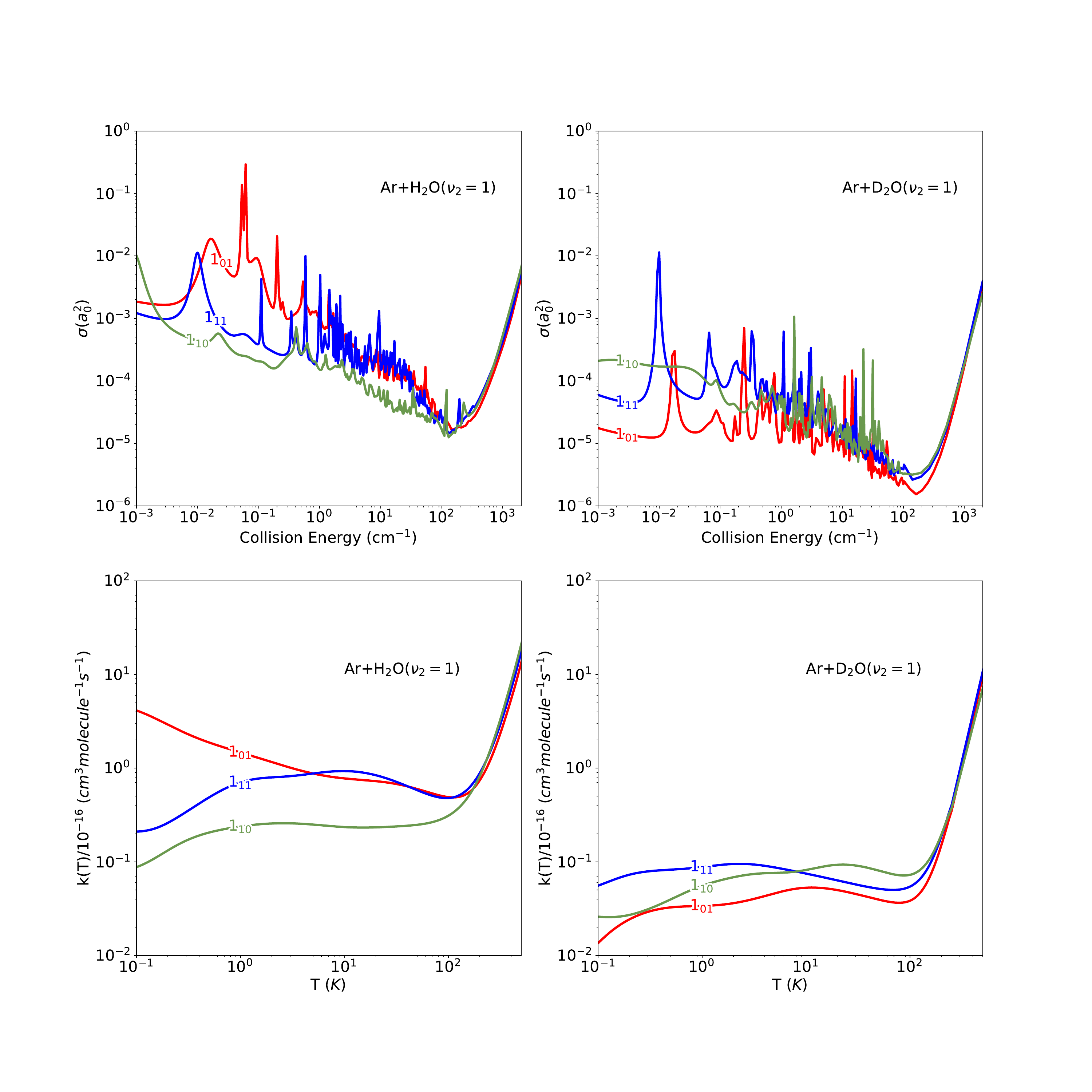}
	\caption{Top panels show the vibrational quenching cross sections for the \ce{H2O} (left) and \ce{D2O} levels (right) belonging to the $j=1$ multiplet. Bottom panels presents the vibrational quenching rate coefficients for the same species and the same levels.}
	\label{vibquenj1}
\end{figure*}

The vibrational quenching cross sections and rate coefficients of the three ($\nu_{2}=1,1_{k_{A}k_{C}}$) levels of \ce{H2O} and \ce{D2O} are presented respectively in the left and right panels of Fig. \ref{vibquenj1}. The cross sections  follow the same variation patterns as a function of collision energy  than those analysed for the collisions of water with \textit{para}-H$_2$($j=0$)\cite{Stoecklin-rb-2019}, He\cite{Stoecklin-H2o-He-2021} and H \cite{lisan-h-h2o}. As noticed in these previous works, the cross section increase monotonously for collision energies higher than the PES well depth  and can be efficiently  fitted to an energy power law allowing the extrapolation  to collision energies higher than those calculated.  The resulting temperature dependence of the state selected rates presented in the lower panels of Fig. \ref{vibquenj1} depends on the initial rotational level for temperatures in the [0.1,100.0] K, while for temperatures higher than 100 K approximately associated with the PES well depth, it is almost identical for the three states ($1_{01}$, $1_{11}$ and $1_{10}$) . Strong differences between the collisions with \ce{H2O} and \ce{D2O} are observed in the [0.1,100.0] K interval where the $1_{10}$ rotational state gives the lowest rate for the collisions with \ce{H2O} while for D$_2$O it is the  $1_{01}$ level. Collisions with water give larger quenching rates than those with D$_2$O for almost the entire temperature interval represented. Nevertheless, the differences between the \ce{H2O} and \ce{D2O} rates becomes smaller at higher temperature, where the rates for the two molecules are of the same order of magnitude. At 500 K, the ratios of the state selected \ce{H2O}  and D$_2$O rate remain greater than 1 being respectively equal to 1.4, 1.6 and 2.9 for the $1_{01}$, $1_{11}$ and $1_{10}$ states.

\begin{figure}[!ht]
	\centering
	\includegraphics[width=\textwidth]{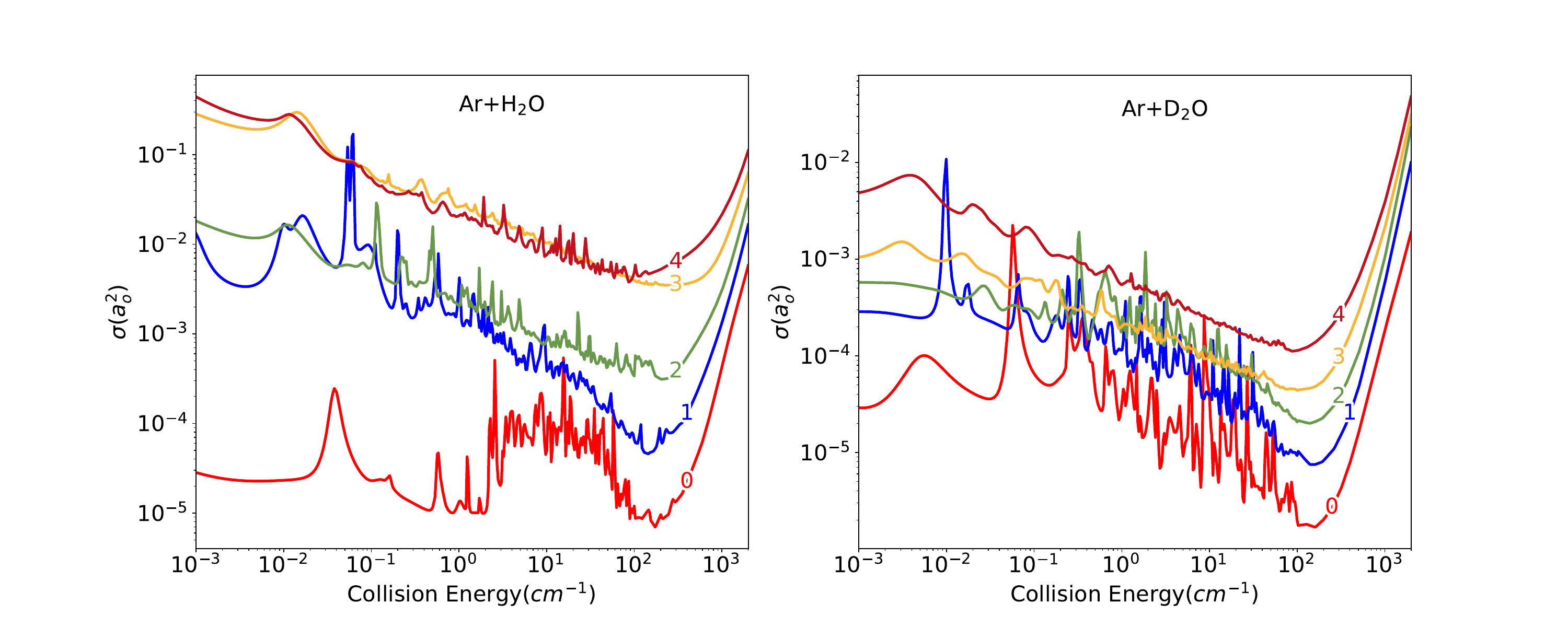}
	\caption{Vibrational quenching cross sections for all the contributions of the rotational states inside a given $j$-multiplet as a function of the collisional energy. Upper and Bottom panels are used for \ce{H2O} and \ce{D2O} respectively.}
	\label{vibquen}
\end{figure}

The $j$ selected vibrational quenching cross sections summed over all the  contributions inside a given $j$-multiplet are presented as a function of  collision energy in the left and right panels of Fig. \ref{vibquen} respectively for the collisions of \ce{H2O} and \ce{D2O} with Ar. The bending cross section is seen to increase monotonously  as a function of the initial \ce{H2O}/\ce{D2O} rotational excitation demonstrating  the strong coupling between bending and rotation.  The  $j=4$ cross section is for example seen to be several orders of magnitude larger than the $j=0$ one. 

\subsubsection*{Comparison with Experiment}

\begin{figure}[!ht]
	\centering
	\includegraphics[width=\textwidth]{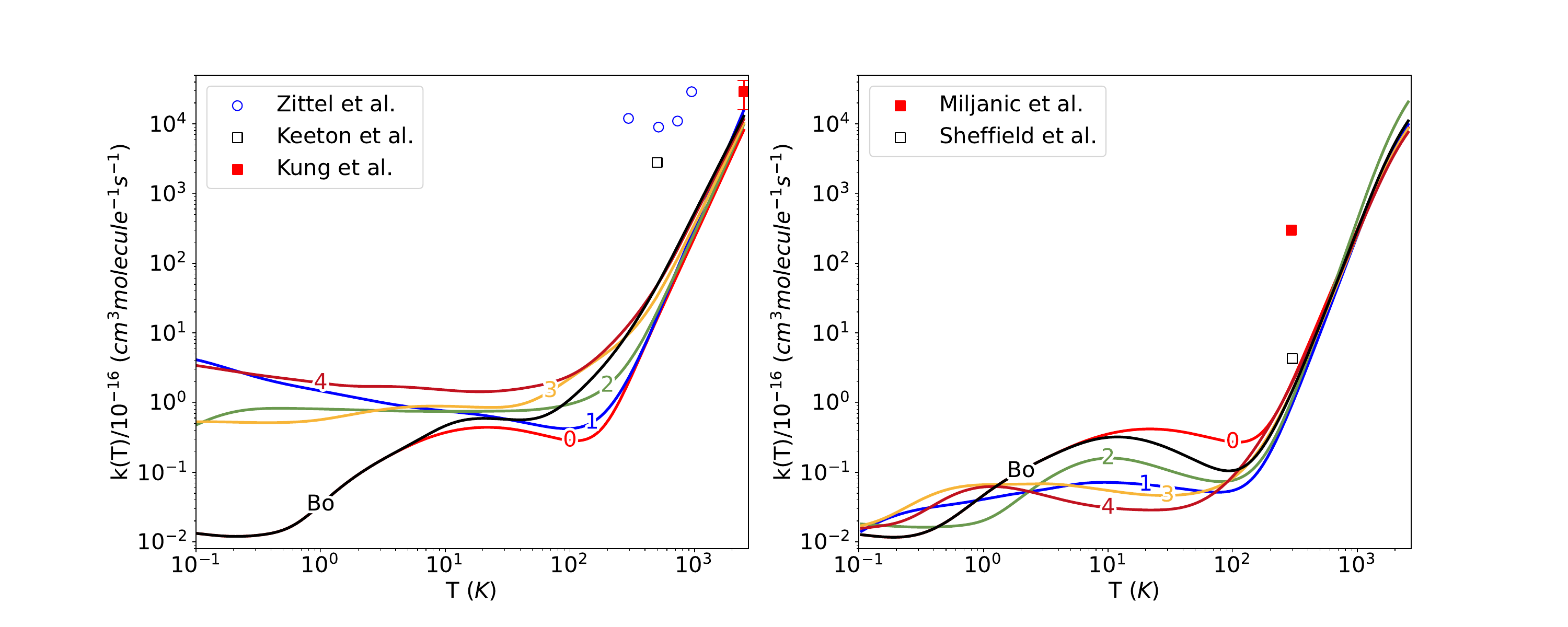}
	\caption{Comparison between the state selected ($\nu_2 = 1, j=0,1,2,3,4 $) bending relaxation rates  of \ce{H2O} (upper panel) and \ce{D2O} (bottom panel)   by collision with Ar. The global Boltzmann averaged rates (Bo) are also reported as well as the experimental measurements given in Refs. (\citenum{Zittel1989,keeton1976,kung1975high,Sheffield1980,Miljanic1980}).} 
	\label{rates}
\end{figure}

Experimental measurements of the bending relaxation  rate coefficient are available for both systems. The calculated $j$-selected rates presented in Fig. \ref{rates} were  then Boltzmann-averaged to obtain the global one which are compared to the experimental values. The rates in the [500,2500] K interval were obtained from the extrapolated energy power law of the cross sections mentioned in the previous section, the coefficients for every rotational initial state can be found in the Suplementary File. The results for \ce{H2O} and \ce{D2O} are presented in the left and right panels of Fig. \ref{rates}, respectively. For both systems the two previously discussed different regimes are clearly identified, one at low temperature where the temperature dependence of the $j$-selected rates is different for all the initial rotational states and a second one for intermediate and high temperatures showing a monotonous increase as temperature increases. \\

Two sets of experimental data are available for the collisions of \ce{H2O} with Ar. Our rates are consistent with the upper limits given by the  Zittel and Masturzo team \cite{Zittel1989} but are smaller than the  estimate reported by Keeton and Bass \cite{keeton1976} at 500K by approximately two orders of magnitude. In this older work the authors however specify that the vibrational relaxation time for \ce{H2O} + Ar, from which the rates are calculated, can only be considered accurate to within an order of magnitude. The discrepancy between our estimate and the one of this experimental team  can also be due to the model they used for the calculation of the classical absorption and  to the extrapolation they did in the 0\% \ce{H2O} mixture limit. At 2500 K the agreement between the rate reported by Kung and
Center \cite{kung1975high} and our theoretical results is quite satisfactory. As a matter of fact, the experimental value reported is $(2.90 )\times 10^{-12}$ cm$^{3}$molecule$^{-1}$s$^{-1}$ and its considered by the authors to be accurate within a factor of 2, while our calculated value for the Boltzmann averaged rate is $1.14 \times 10^{-12}$ cm$^{3}$molecule$^{-1}$s$^{-1}$, which is very close to the predicted experimental limits.   \\         

Two contradicting experimental values were reported the same year for the collisions of Ar with D$_2$O at about 295 K.  Miljanić and Moore \cite{Miljanic1980} reports a value of $3.0 \times 10^{-14}$ cm$^{3}$molecule$^{-1}$s$^{-1}$ while the one obtained from the vibrational relaxation time reported by Sheffield \etl \cite{Sheffield1980} is, as was mentioned in Ref~\citenum{Miljanic1980}, about 70 times smaller $\approx 4.29 \times 10^{-16}$ cm$^{3}$molecule$^{-1}$s$^{-1}$. Our theoretical value  ($1.42 \times 10^{-16}$ cm$^{3}$molecule$^{-1}$s$^{-1}$ )  is then closer to the one of Sheffield \etl (about three times smaller ).  The comparison between theory and experiment for these two systems suggests that up to date  experimental data are badly needed for these two systems.

%\subsection{Ar-D$_2$O system}

\section*{Conclusion}\label{Concl}
The bending relaxation of \ce{H2O} and \ce{D2O} by collisions with Ar was studied at the Close Coupling level taking fully into account the coupling between bending and rotation within the RB-CC method. To this aim two new 4D PES were developed which accuracy was tested at the rigid rotor level by performing RR-CC scattering calculations as well as computing  the \ce{D2O-Ar} bound states and comparing the results with available theoretical and experimental data. Both the dynamics and the bound states results were found to be in excellent agreement with all these data.  The propensity rules we obtained previously for the rotationally inelastic collisions of \ce{H2O} with  Ne or He were found to hold for \ce{H2O} + Ar. We conversely could not infer any convincing propensity rules for the collisions of \ce{D2O} with Ar except for the most excited  level inside any $j$ multiplet which follows the  rules obtained for the collisions with \ce{H2O}. However, the state to state cross sections  in Ne + D$_2$O and Ar + D$_2$O seem to follow the same ordering suggesting that, some more complicated rules associated with the smaller energy gaps in D$_2$O are at play and still need to be inferred. We then computed the bending relaxation cross sections and rates for these two systems. We found that bending relaxation is more efficient when the rotation of \ce{H2O} or \ce{D2O} is excited suggesting that bending and rotation are coupled. We also found bending relaxation to be more efficient with \ce{H2O} than \ce{D2O}. The larger spacing between the rotational states of \ce{H2O} may explain this result by increasing the coupling between the rotational states belonging to the fundamental and excited bending level of \ce{H2O}. We eventually compared the bending relaxation rates with their experimental estimates. We obtain a reasonable agreement with some sets of  experimental data but some other sets differ from them making difficult to conclude about the validity of the comparison. New up to date experiments are  then needed for these two systems.

%\end{linenumbers}
\end{doublespacing}

\section*{Acknowledgments}
This work was supported by the french Agence Nationale de la Recherche (ANR-Waterstars), Contract No. ANR-20-CE31-0011.  We acknowledge the support from the  ECOS-SUD-CONICYT project number C22E02 (Programa de Cooperaci\'on Cient\'ifica ECOS-ANID ECOS220023). LDCH  also thanks the support from the  ANID BECAS/DOCTORADO NACIONAL 21210379. Computer time for this study was provided by the M\'{e}socentre de Calcul Intensif Aquitain, which is the computing facility of Universit\'{e} de Bordeaux et Universit\'{e} de Pau et des Pays de l'Adour.

\section*{Conflicts of interest}
There is no conflict of interest to report.
%%%%%%%%%%%%%%%%%%%%%%%%%%%%%%%%%%%%%%%%%%%%%%%%%%%%%%%%%%%%%%%%%%%%%
\renewcommand\refname{References}
%%%%%%%%%%%%%%%%%%%%%%%%%%%%%%%%%%%%%%%%%%%%%%%%%%%%%%%%%%%%%%%%%%%%%
\begin{footnotesize}
\bibliographystyle{unsrt.bst} % abbrvnat or unsrt
\textnormal{\bibliography{main}}
\end{footnotesize}
\newpage

%%%%%%%%%%%%%%%%%%%%%%%%%%%%%%%%%%%%%%%%%%%%%%%%%%%%%%%%%%%%%%%%%%%%%
% Tables
%%%%%%%%%%%%%%%%%%%%%%%%%%%%%%%%%%%%%%%%%%%%%%%%%%%%%%%%%%%%%%%%%%%%%

\end{document}